\documentclass[aip,jap,reprint]{revtex4-1}
\usepackage[utf8]{inputenx} 
\usepackage[english]{babel} 
\usepackage{subfigure} 
\usepackage{hyperref} 
\usepackage{graphicx}
\usepackage{amsmath,amssymb}

\newcommand{\PO}{$^{210}$Po}
\newcommand{\TEO}{$\mathrm{TeO}_2$}
\newcommand{\TEHT}{$^{130}\mathrm{Te}$}

\newcommand{\Ohm}{\un\Omega}
\newcommand{\MO}{\un{M\Ohm}}
\newcommand{\GO}{\un{G\Ohm}}
\providecommand*{\un}[1]{\ensuremath{\mathrm{\,#1}}}

\begin{document}

\title{Model of the response function of large mass bolometric detectors}
\author{M.~Vignati}
\email[]{marco.vignati@roma1.infn.it}
\affiliation{Dipartimento di Fisica, Sapienza Universit\`a di Roma and Sezione INFN di Roma, I-00185, Italy}
\date{\today}
\begin{abstract}
Large mass bolometers are used in particle physics experiments to search for rare processes.
By operating at low temperature, they are able to detect particle energies 
from few keV up to several MeV, measuring the temperature rise produced by the energy released.
This study was performed on the bolometers of the CUORE experiment.
The response function of these detectors is not linear in the energy range of interest,
and it changes with the operating temperature.
The non-linearity is found to be dominated by the thermistor and its biasing circuit, and is modeled
using few measurable parameters.
A method to obtain a linear response is the result of this work. It allows a great simplification of the
data analysis.
\end{abstract}

\maketitle

\section{Introduction}
Bolometers are detectors where the energy of particle interactions is converted into
heat  and measured via temperature variation. They provide
an excellent energy resolution, although they are slow compared to conventional detectors.
These features make them a suitable choice for experiments looking to
rare processes, like neutrinoless double beta decay and dark matter interactions.

The CUORE experiment searches for neutrinoless double beta decay in \TEHT~\cite{Ardito:2005ar}, 
using bolometers made of \TEO\ crystals whose temperature is measured by neutron transmutation
doped (NTD) Germanium semiconductor thermistors~\cite{wang,Itoh}.
Operated at temperatures of about $10\un{mK}$,
these detectors maintain a resolution of few keV over their energy range, 
extending from few keV up to several MeV.
In this range the response function is found to be non-linear. 
The conversion from signal amplitude to energy is complicated and the shape of the signal 
depends on the energy itself.
Moreover the amplitude of the signal depends on the temperature of the detector,
which cannot be kept stable by current cryostats at a level that would not perturb the resolution.
Understanding the non-linearities is important to have a reliable calibration,
to control the signal shape and to improve the resolution.

Electrothermal models for large mass bolometers have been already proposed~\cite{JONES:53,pessinamod,McCammon200411}.
Based on  systems of differential equations, they describe the 
behavior of the detector and reproduce the shape of the signal 
and some features of the calibration function. However, they rely on many parameters 
that are difficult to measure with the necessary precision in working conditions. These models are
useful to describe the response of a bolometer, but they cannot produce tools
to be used on a real detector.

In this paper a model based on few parameters is proposed.  It shows that the non-linearities 
are dominated by the thermal sensor and by the biasing circuit that is used to read it.
These elements are fully measurable in working conditions, so that the model can be
applied to the detector.
A method to remove the non-linearities is derived. Its application in the data analysis
linearizes the response function, improving the results.
\section{Description of the detector}
\begin{figure}[htbp]
\begin{center}
\includegraphics[clip=true,width=0.4\textwidth]{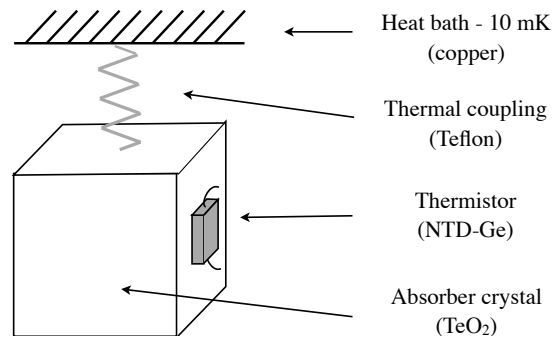}
\caption{Sketch of a CUORE bolometer. The \TEO\ crystal is held by Teflon
supports, the thermistor is glued to the crystal and its wires are glued to copper frames. The supports and the thermistor wires thermally couple the crystal to the copper frames, which act as heat bath. }
\label{fig:crystal_setup}
\end{center}
\end{figure}

A CUORE bolometer is composed of two main parts, a \TEO\ crystal and
a NTD-Ge thermistor.
The crystal is cube-shaped, weighs 750\un{g} and its heat capacitance $C$ at the working temperature of 10\un{mK} is of order 2\un{nJ/K}~\cite{Barucci}.
It is held by Teflon supports on copper frames, which are connected to the mixing chamber
of a dilution refrigerator. The thermistor is glued to the crystal and the biasing wires are glued 
to the copper frames (see Fig.~\ref{fig:crystal_setup}) .

When an amount of energy $E$ is released in the crystal, its temperature increases by $E/C$ and then returns to equilibrium via the thermal conductance of the supports and of the thermistor. 
The latter acts as thermometer, converting temperature, $T$, into resistance, $R$,
according to the relationship~\cite{Mott:1969}:
\begin{equation}
R(T) = R_0\exp \left(T_0 / T \right)^\gamma
\label{eq:thermistor}
\end{equation}
where $R_0$ and $T_0$ are parameters that depend on the dimensions and on the material of the thermistor.
Their value is about $1.1\un{\Ohm}$ and $3.4\un{K}$ respectively.
At 10\un{mK} the parameter $\gamma$ can be considered constant and equal to $1/2$ ~\cite{efros,Itoh:1996}.

\begin{figure}[htbp]
\begin{center}
\includegraphics[clip=true,width=0.4\textwidth]{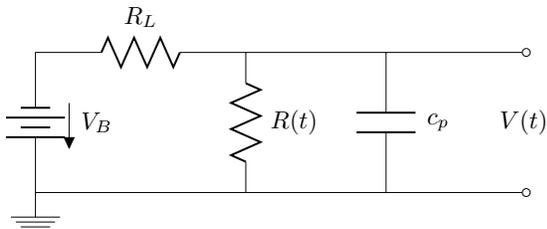}
\caption{Biasing circuit of the thermistor. A voltage generator $V_B$ biases the thermistor resistance $R(t)$ in series with a load resistance $R_L$. The bolometer signal is the voltage $V(t)$ across $R(t)$. The wires used to read $V(t)$ have a non-negligible capacitance $c_p$.
}
\label{fig:biasing}

\end{center}
\end{figure}

To read out the signal, the thermistor is biased with constant current, which is provided by
a voltage generator, $V_B$, and a load resistor, $R_L$, in series with the thermistor. 
The resistance of the thermistor varies in time with the temperature, $R(t)$, and the voltage across it, $V(t)$,
is the bolometer signal. 
The value of $R_L$ is chosen to be much higher than $R(t)$ so that  $V(t)$ is almost proportional to $R(t)$.
Since from Eq.~(\ref{eq:thermistor}) positive temperature variations induce negative resistance variations,
the voltage $V_B$ is chosen to be negative in order to obtain positive signals.  
The connecting wires add in parallel to the thermistor a parasitic  capacitance $c_p$ (see Fig.~\ref{fig:biasing}). The signal $V(t)$ is amplified, filtered with a 6-pole active Bessel filter and then acquired
with an 18 bit DAQ system. The front end electronics, which provide the bias, the load resistors, and the amplifier, are placed outside the cryostat, at room temperature~\cite{AProgFE}.

At 10\un{mK} the value of $R(T)$ is of order $100\MO$, $R_L$
is chosen as $\sim 50\GO$ and $V_B$ as $\sim 5$\un{V}. 
The $c_p$ value depends on the length of the wires that carry the signal out of the cryostat, typically it is of order 400\un{pF}. The amplifier gain, the Bessel frequency bandwidth, the duration of the
acquisition window and the sampling frequency are set typically at 5000\un{V/V}, 12\un{Hz},
5.008\un{s} and 125\un{Hz} respectively.

\begin{figure}[htbp]
\begin{center}
\includegraphics[width=0.48\textwidth]{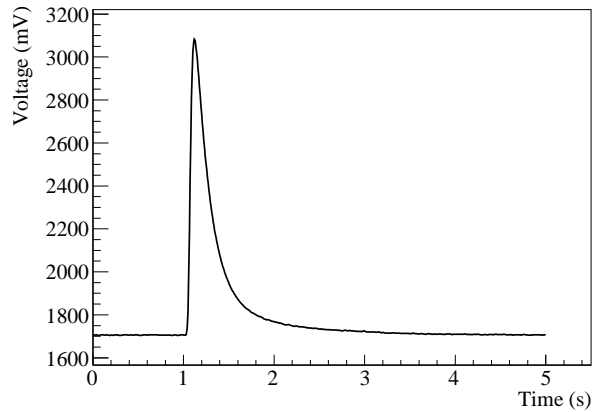}
\caption{Pulse generated by a 2615\un{keV} $\gamma$ particle. The baseline is related to the temperature
of the thermistor before the particle interaction, the amplitude carries information on the amount of energy released.}
\label{fig:pulse}
\end{center}
\end{figure}

A typical signal acquired by the ADC, is shown in Fig.~\ref{fig:pulse}. 
The signal was generated by a  2615\un{keV} $\gamma$ particle fully absorbed in the crystal.
The baseline voltage of the pulse is related to the thermistor temperature before the particle interaction, and
the amplitude is related to the energy released.

\section{Non-linearities of the response function}
\begin{figure}[htbp]
\begin{center}
\includegraphics[width=0.48\textwidth]{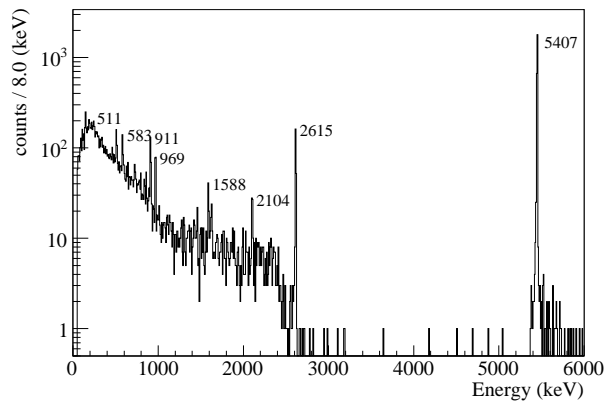}
\caption{Energy Spectrum. The $\alpha$ line at 5407\un{keV} is generated by \PO\ contamination of the crystal,
the rest of the spectrum is due to $\gamma$'s and $\beta$'s generated by a $^{232}$Th source. The most prominent peaks are labeled with their energy (in keV).}
\label{fig:energy_spectrum}
\end{center}
\end{figure}

The data analyzed in this paper have been obtained by exposing the bolometer to a $^{232}$Th source. This source, 
together with an $\alpha$ line generated by \PO\ contamination of the crystal,
allows the analysis of an energy range up to 5407\un{keV} (see the energy spectrum in 
Fig.~\ref{fig:energy_spectrum}).

In this range several non-linearities of the response function are observed:
 \begin{enumerate}
 \item {
The shape of the pulse is not constant with energy. The rise and the decay time, 
computed as the time difference between the 10\% and the 90\% of the trailing edge and the 90\% and 30\%
of the leading edge respectively, are correlated with energy (see Fig.~\ref{fig:shape_std}). 
This effect makes any pulse shape discrimination in the data analysis difficult.
\begin{figure}[htbp]
\centering
\subfigure[]{
    \includegraphics[width=0.48\textwidth]{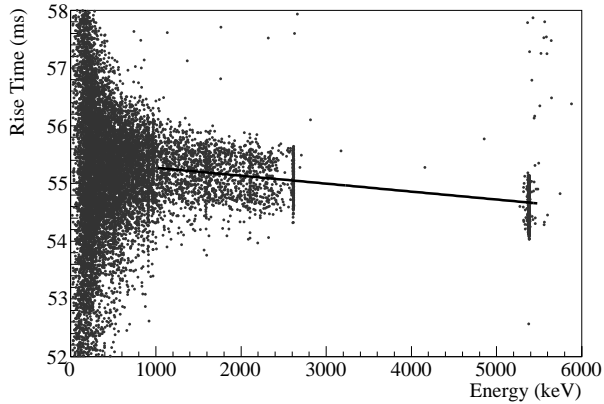}  
 }
 \subfigure[]{
    \includegraphics[width=0.48\textwidth]{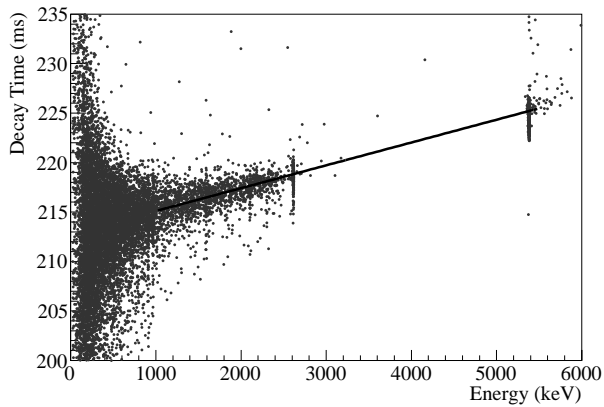}  
  }
\caption{Pulse shape parameters versus energy. The rise time (a) and the decay time (b) are anti-correlated and 
correlated with energy respectively.}
\label{fig:shape_std}			
\end{figure}
}
\item {
The amplitude of the pulse depends on the base temperature. In figure \ref{fig:stab_std} the amplitude of
5407\un{keV} events is plotted with respect to the pulse baseline, showing a clear anti-correlation.
The temperature variation of the baseline is about $40\un{\mu K}$, which corresponds
to an energy variation of the crystal of about $400\un{keV}$. The pulse amplitude changes by about $1 \%$,
worsening the resolution of the detector. This effect
can be partially corrected by applying a calibrated heat source to the detector~\cite{stabilization,Arnaboldi:2003yp} and
stabilizing the base temperature~\cite{Arnaboldi:2005xu}.
\begin{figure}[htbp]
\begin{center}
\includegraphics[width=0.48\textwidth]{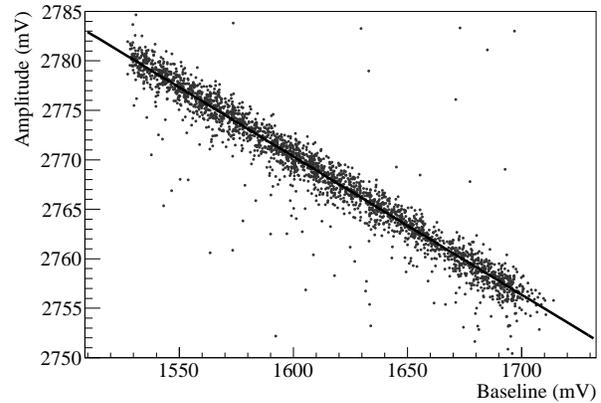}
\caption{Amplitude of 5407\un{keV} pulses versus baseline. A change in the bolometer temperature
also changes its response, worsening the resolution.}
\label{fig:stab_std}
\end{center}
\end{figure}
}
\item {
The amplitude dependence on energy is not linear, and is usually parameterized with a polynomial function.
The deviation from linearity of the data is estimated using a linear calibration function 
\begin{equation}
\rm{Energy} = \rm{constant}\cdot\rm{Amplitude}
\end{equation}
and then the difference between the energy estimated with this function and the nominal one is computed.  
The residuals evaluated on the peaks labeled in Fig.~\ref{fig:energy_spectrum}  
are shown in Fig.~\ref{fig:res_std}. The 5407\un{keV} line is not
included, since $\alpha$ particles have a quenching factor different from $\gamma$ and $\beta$
particles~\cite{alessandrello:1998na}. 
If the true calibration function were a line, the residuals would be much smaller
than the energy resolution, that is of order 5\un{keV} FWHM. The data, however, clearly indicate
that a line is not sufficient to describe the calibration function.
\begin{figure}[htbp]
\begin{center}
\includegraphics[width=0.48\textwidth]{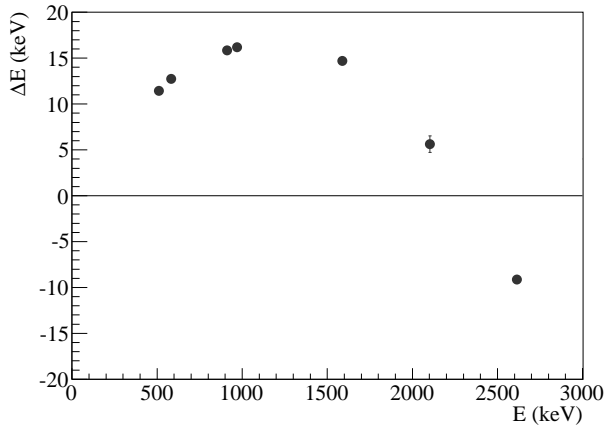}
\caption{Residuals obtained using a linear calibration function. The difference $\Delta E$  between the
estimated energy and the nominal one is not compatible with zero, considering that  the energy resolution is $\sim5\un{keV\,FWHM}$. The error bars refer to the uncertainty on the estimated peak position, that depends
on the FWHM and on the number of events $N$ in the peak as ${\rm FWHM} / (2.35 \sqrt{N})$.}
\label{fig:res_std}
\end{center}
\end{figure}
}
\end{enumerate}

\section{Thermistor and biasing circuit models}
The detector contains many elements that can generate the observed non-linearities. The heat capacitance and
conductance of the crystal, of the Teflon supports and of the thermistor, all depend on the temperature. 
The crystal capacitance for example is not linear in temperature, as it follows the Debye law ($C \propto T^{3}$). 
The problem of modeling the response function is that not all elements
can be measured with satisfactory precision. 
In particular the contact heat conductance between the crystal and its supports
changes with the detector configuration and the heat conductance of the glue
varies from bolometer to bolometer because of the weak reproducibility of the glue deposition.
In principle the knowledge of all parameters is needed if we want to predict the response function. 
However, we can narrow our focus to the aspects of the physical system which produce
the dominant contributions to the non-linearity. 
In this section we will develop a model of the thermistor response and of the biasing circuit.
It will be shown that they can account for all the observable non-linearities, 
without using any thermal model. 

First we have to extract the temperature of the thermistor from the measured resistance.
Unfortunately we cannot use Eq.~(\ref{eq:thermistor}), since parameters $R_0$ and $T_0$ cannot be measured with 
precision. This is due to the fact that when the thermistor is attached to the crystal,
the glue stresses the device modifying its features in an unpredictable way. 
However, since we are interested in measuring energy depositions and not absolute  temperatures,
we can just evaluate the temperature variation. The resistance variation, $\Delta R$, in terms of the temperature variation, $\Delta T$, can be approximated as:
\begin{align}
\label{eq:thermistor_approx}
\nonumber
\Delta R(\Delta T) &= R(T+\Delta T) - R(T) \\
                                &\simeq R(T) \left[ \exp(-\eta \Delta T / T ) - 1\right]
\end{align}
where:
\begin{equation}
\eta = \left|\frac{d \log{R}}{d \log T} \right|  = {\gamma} \log\frac{R(T)}{R_0}
\end{equation}
is the sensitivity of the thermistor and its value is of order 10. 
The reason for expanding  the exponential argument in terms of $\Delta T/T$ instead of the entire formula
is that it provides a better approximation.
The approximation is valid for $\Delta T/T \ll 1$, or, in
terms of particle energy, for $E \ll C\,T$. Since $C\,T$ is of order $100\un{MeV}$ ($C$ and $T$ are about 10\un{MeV/mK} and 10\un{mK} respectively), the approximation holds in the entire energy range
\footnote{We assumed that the temperature increase of the thermistor is equal to that of the crystal.
Since there is a thermal coupling between them, the increase is lower, making the approximation even more valid.} 
.
The advantage of Eq.~(\ref{eq:thermistor_approx}) is that it allows $\Delta T$ to be extracted from $\Delta R$
using a measurable parameter (the resistance $R(T)$) and that the unmeasurable parameter ($\eta/T$)
is just a  scale factor. In order to simplify the notation we define $\sigma = \eta /T$ for future calculations.

In the next section we will put in evidence the non-linearities generated by Eq.~(\ref{eq:thermistor_approx}) 
making the following assumptions:
\begin{enumerate}
\item The temperature response of the detector after an energy deposition $E$ can be parametrized as 
\begin{equation}
\Delta T(t,E) = A(E)\,U(t)
\end{equation}
 where $U(t)$ is a function representing the time evolution of the temperature pulse and $A(E)$ is 
 the amplitude of the pulse, which depends on the energy deposited.
\item The pulse shape $U(t)$ is independent of the energy and has unitary amplitude.
\item The amplitude $A(E)$ is proportional to the energy.
\item Both $A$ and $U$ are independent of the baseline temperature (the temperature before the energy deposition).
\end{enumerate}
The properties of the resistance signal, $\Delta R(t)$, obtained under these assumptions will be compared
with the measured voltage signal, assuming that the biasing circuit does not add non-linearities. 

The dependence of the shape parameters on energy (see Fig.~\ref{fig:shape_std}) is obtained computing the time derivative of the signal $\Delta R(t,E)$ normalized  to its amplitude $A_R(E)$: 
\begin{align}
\nonumber
\frac{\Delta R'(t,E)} {A_R(E)} &= -
 \frac{ \exp[-\sigma A(E)\, U(t)]\, \sigma\, A(E)\, U'(t)} {\exp[-\sigma A(E) ] - 1}\\
  &\simeq U'(t) \left[1-\sigma A(E)\, U(t)\right]\,,
\end{align}
where $A_R(E) = \min_t[\Delta R(t,E)]$ (since $\Delta R(t,E)$ is a negative signal) and $U'(t)$ is the time derivative of $U(t)$. This expression indicates that the signal is slower when the energy increases, the same effect seen
on the decay time of the pulse. This equation does not explain the behavior
of the rise time, that is dominated by the biasing circuit (see later in the text).

The dependence of the pulse amplitude on baseline is also straightforward to derive. 
The resistance variation of the pulse baseline depends on the temperature according to the expression:
\begin{equation}
\Delta R _B = R(T) \left[ \exp(-\sigma \Delta T_B ) - 1\right]
\end{equation}
where $\Delta T_B$ is the temperature variation with respect to $T$. When energy is released, the temperature increases by $A(E)$, and the total resistance change will be: 
\begin{equation}
\Delta R_{Tot}  = R(T) \left[ \exp(-\sigma (\Delta T_B+A(E))) - 1\right].
\end{equation}
The amplitude of the pulse, however, is computed with respect to the baseline. In terms of resistance variation it will be:
\begin{align}
\nonumber
A_R(E,\Delta R_B) &= \Delta R_{Tot} - \Delta R_B \\
&\simeq \left[1+\frac{\Delta R_B}{R(T)}\right] A_R(E,\Delta R_B = 0)\, .
\end{align}
Recalling that a baseline increase in Volts corresponds to a resistance decrease, this expression
describes the anti-correlation of the pulse amplitude with its baseline (see Fig.~\ref{fig:stab_std}). 

 The non-linearity of the calibration function comes directly from Eq.~(\ref{eq:thermistor_approx}).
If we assume that $A(E) =  E/C$, the pulse amplitude $A_R(E)$ at the second order increases with energy as:
\begin{equation}
A_R(E) = R(T) \left[ -\sigma\frac{E}{C} +  \left(\sigma\frac{E}{C}\right)^2 + \ldots\right]\, .
\end{equation}
Recalling that the signal in Volts has opposite sign of the resistance variation, this is the same behavior of 
Fig.~\ref{fig:res_std}.

The thermistor formula (\ref{eq:thermistor_approx})
 explains at this point at least qualitatively all the data, except for the rise time of the pulse.
The calculations showed in fact that the shape parameters should get slower with amplitude.
The rise time, however, becomes faster (see Fig.~\ref{fig:shape_std}). 
This effect can be generated by the biasing circuit,
whose parasitic capacitance, $c_p$, introduces a cutoff that varies with the pulse amplitude.
We solve the circuit of Fig.~\ref{fig:biasing} approximately, using $R_L \gg R$ and $R_L\gg \Delta R(t)$. 
We have that the voltage signal $\Delta V(t)$
in terms of the resistance signal $\Delta R(t)$ can be expressed as:
\begin{equation}
\Delta V(t) + \left[R(T) + \Delta R(t)\right] c_p  \Delta V'(t) = -\frac{V_B}{R_L}\Delta R(t), 
\label{eq:cutoff}
\end{equation}
where $\Delta V(t)= V(t) - V_R$, and $V_R$ is the voltage corresponding
to the static resistance $R(T)$: $V_R = -V_B \,R(T) / [R(T)+R_L]$.
Expression (\ref{eq:cutoff}) has the form of a low pass filter whose cutoff, $[c_p (R(T)+\Delta R(t))]^{-1}$, depends on  $\Delta R(t)$ and hence on its amplitude. The higher
is the absolute value of the amplitude, the higher is the cutoff (since the amplitude is negative), making the voltage pulse faster.
Since $R(T)c_p$ is of order 30\un{ms}, this effect is visible only on the fast part of the voltage pulse (the rise time)
and competes with the thermistor slowing down.

\section{Linearization of the response function}
It has been shown that the behavior of the data can be explained assuming that the 
thermal response $\Delta T(t)$ does not depend on the energy released or on the variation of the baseline
bolometer temperature. The non-linearities of the voltage 
signal seem to be generated by the thermistor and by the biasing circuit.
Here we derive an algorithm to extract a signal proportional to $\Delta T(t)$,
and we will test its performance on data.

\begin{figure}[htbp]
\centering
    \includegraphics[width=0.48\textwidth]{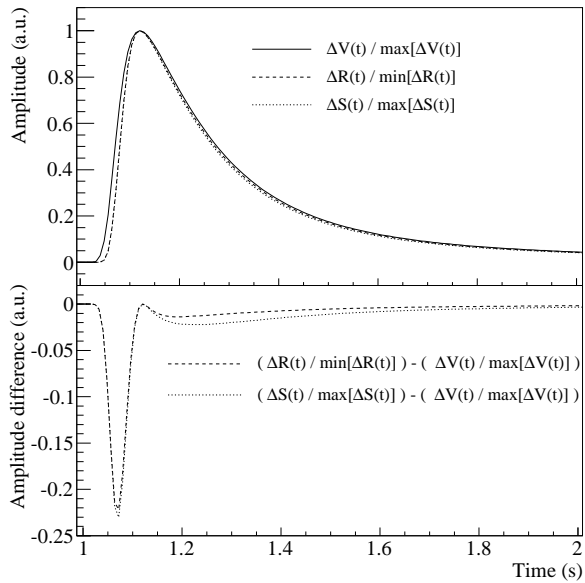}  
\caption{Comparison between the original $\Delta V(t)$ signal generated by a 2615 keV $\gamma$ particle, and the same signal transformed into $\Delta R(t)$ and $\Delta S(t)$. The $\Delta R(t)$ signal is faster than the $\Delta V(t)$ one, because the low pass filter of the biasing circuit has been deconvoluted. The $\Delta S(t)$ signal is even faster
since the thermistor slowing down effect has been removed.}
\label{fig:shape_compare}			
\end{figure}

First the variation of the thermistor resistance is extracted from $V(t)$ solving the biasing circuit of 
Fig.~\ref{fig:biasing}:
\begin{equation}
\Delta R(t) = -
\frac
{\Delta V(t) [R(T)+R_L] + R_L\,R(T)\,c_p  \Delta V' (t)}
{\Delta V(t) + R_L c_p \Delta V'(t) + V_B\, R_L/[R(T) + R_L]}
\,,
\end{equation}
where $\Delta V(t)$  is defined in the previous section.
Then from Eq.~(\ref{eq:thermistor_approx})  the quantity $\Delta S(t) = \sigma\,\Delta T(t) $ is computed:
\begin{equation}
\Delta S(t) =-\log\left[1+\frac{\Delta R(t)}{R(T)}\right]\,.
\end{equation}
The advantage of this algorithm is that it depends on parameters that can be measured with sufficient precision.
These parameters are the static bolometer resistance $R(T)$ and the voltage across it $V_R$, the load resistance $R_L$, the parasitic capacitance $c_p$ and the bias voltage $V_B$. 
The unknown quantities have been hidden in a common scale factor $(\sigma)$, that is not relevant
if we are only interested in removing the non-linearities.
This algorithm has been implemented as a digital filter on the voltage data sampled by the ADC. The derivative of 
$\Delta V$ has been computed as $\Delta V'_i = (V_{i+1}-V_{i-1})/(2\,\Delta t)$, 
where $i$ is the sample index and $\Delta t$ is the sampling period. 
A comparison between the original $\Delta V(t)$ signal and the same signal transformed into $\Delta R(t)$ and $\Delta S(t)$ is shown in Fig.~\ref{fig:shape_compare}.

The $\Delta S$ data have been analyzed with the same analysis procedure applied on $\Delta V$ data. The features
of the transformed signal are:
\begin{enumerate}
\item {The shape of the pulse has a much weaker dependence on energy. The correction is particularly evident on the decay time, where the correlation has been reduced by a factor 6 (see Fig.~\ref{fig:shape_std_tr}).
\begin{figure}[htbp]
\centering
\subfigure[]{
    \includegraphics[width=0.48\textwidth]{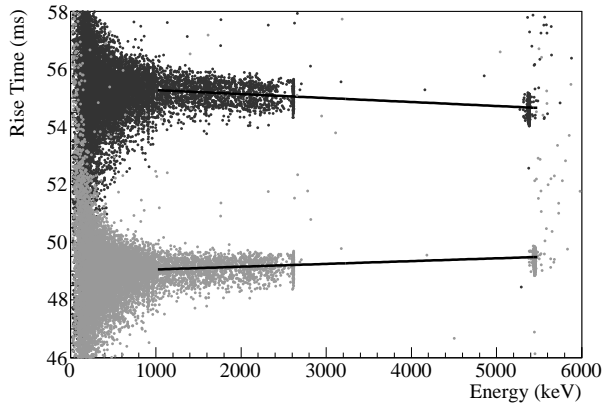}  
 }
\subfigure[]{
    \includegraphics[width=0.48\textwidth]{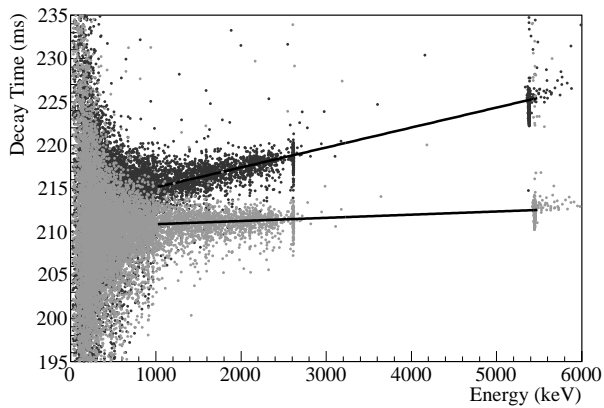}  
 }
\caption{Pulse shape parameters versus energy. The $\Delta S$ data (light gray) are less dependent on energy
than $\Delta V$ data (dark gray), the correction is particularly effective on the decay time (b). 
The rise time (a) of $\Delta S$ data is faster, because the low pass filter of the biasing circuit has been deconvoluted.}
\label{fig:shape_std_tr}			
\end{figure}
}
\item{The amplitude dependence on the baseline is removed. The slope is reduced by a factor 20 
(see Fig.~\ref{fig:stab_std_tr}).
\begin{figure}[htbp]
\begin{center}
\includegraphics[width=0.48\textwidth]{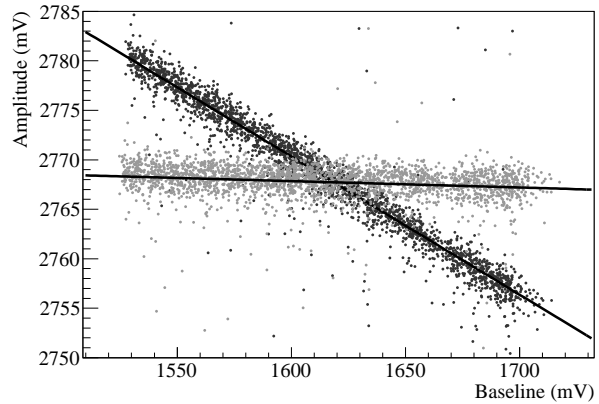}
\caption{Amplitude of 5407\un{keV} pulses versus baseline. The $\Delta S$ data (light gray) are less correlated with
the baseline than $\Delta V$ data (dark gray). The ratio of the slopes is about 20. }
\label{fig:stab_std_tr}
\end{center}
\end{figure}
}
\item{The calibration function is a line. The comparison of the residuals clearly indicates that 
the non-linearities of the calibration function have been removed (see Fig.~\ref{fig:res_std_tr}).
\begin{figure}[htbp]
\begin{center}
\includegraphics[width=0.48\textwidth]{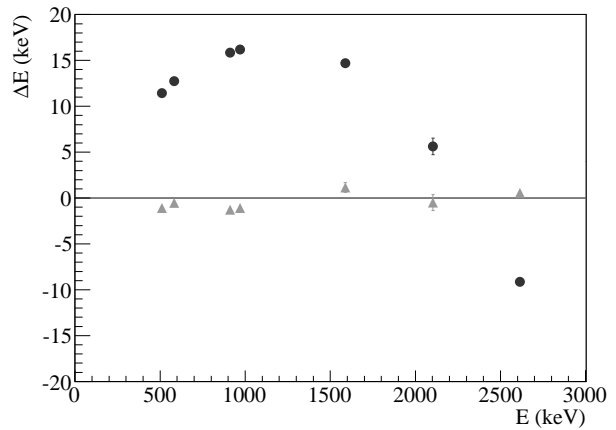}
\caption{Residuals of the linear calibration function. The residuals of $\Delta S$ data (light gray) are very close to zero
and have no trend compared to $\Delta V$ data (dark gray).}
\label{fig:res_std_tr}
\end{center}
\end{figure}
}
\end{enumerate}

In conclusion, the model we have developed describes nearly all the observable non-linearities of the response function. It relies only on fixed parameters, that can be measured on the detector. 
The data transformed with an algorithm derived from the model provide a much more linear
response function, improving the operation and the data analysis of these detectors.

\begin{acknowledgments}
I am very grateful to the members of the CUORE collaboration, in particular to 
C. Bucci, P. Gorla, G. Pessina and S. Pirro for the fruitful discussions
I had with them. I wish to thank F. Bellini and L. Kogler for their valuable suggestions
and F. Ferroni for supporting this work.
\end{acknowledgments}

%

%
\end{document}